\documentclass[aps,reprint,prl,groupedaddress]{revtex4-1}

\usepackage{lineno}
\usepackage[pdftex,
            pdfauthor={Matthias Puhr and Pavel Buividovich},
            pdftitle={A numerical study of non-perturbative corrections to the Chiral Separation Effect in quenched finite-density QCD},
            pdfproducer={Latex with hyperref},
            pdfcreator={pdflatex}]{hyperref}
\usepackage[utf8]{inputenc}
\usepackage[verbose]{placeins}
\usepackage{amsmath}
\usepackage{amssymb}
\usepackage{amsfonts,mathrsfs,fixmath,amsthm}
\usepackage{bbm}
\usepackage{color}
\usepackage{graphicx}
\usepackage{multirow}
\usepackage[caption=false]{subfig}

\newcommand{\Dov}{\ensuremath{D_\text{ov}}}
\newcommand{\K}{\ensuremath{K_{x,\mu}}}

\newcommand{\scsc}{\ensuremath{\sigma_{\text{\tiny CSE}}}}
\newcommand{\scscf}{\ensuremath{\sigma^{\text{\tiny 0}}_{\text{\tiny CSE}}}}

\newcommand{\jxa}{\ensuremath{j_{x,\mu}^5}}

\newcommand{\MeV}{\ensuremath{\text{MeV}}}
\newcommand{\fm}{\ensuremath{\text{fm}}}
\newcommand{\MF}{\ensuremath{\Phi_B}}
\newcommand{\fermi}{\ensuremath{\text{fm}}}
\newcommand{\LS}{\ensuremath{L_S}}
\newcommand{\LT}{\ensuremath{L_T}}
\newcommand{\gpgg}{\ensuremath{ g_{\pi^0\gamma\gamma}}}
\newcommand{\bzero}{\ensuremath{\beta}}

\newcommand{\ket}[1]{\ensuremath{\left|#1\right\rangle}}
\newcommand{\bra}[1]{\ensuremath{\left\langle#1\right|}}
\newcommand{\braket}[2]{\ensuremath{\left\langle#1\right|\left.#2\right\rangle}}

\DeclareMathOperator{\Tr}{tr}
\DeclareMathOperator{\tr}{tr}

\newcommand{\expa}[1]{ \exp{\left( #1 \right)} }

\newcommand{\lr}[1]{ \left( #1 \right) }
\newcommand{\lrs}[1]{ \left[ #1 \right] }

\newcommand{\mycomment}[1]{}

\usepackage{etoolbox}
\newtoggle{arxiv}
\toggletrue{arxiv}

\graphicspath{{/home/pum58515/data/lanczos_calculations/axial_currents/plots/} {./}
  {./plots/}}
\makeatletter
\def\input@path{{./}{./plots/}}
\makeatother

\begin{document}


\title{A numerical study of non-perturbative corrections to the Chiral Separation Effect in quenched finite-density QCD}


\author{Matthias Puhr}
\email[]{Matthias.Puhr@physik.uni-regensburg.de}
\affiliation{Institute of Theoretical Physics, Regensburg University, 93040 Regensburg, Germany}

\author{P.~V.~Buividovich}
\email[]{Pavel.Buividovich@physik.uni-regensburg.de}
\affiliation{Institute of Theoretical Physics, Regensburg University, 93040 Regensburg, Germany}



\date{March 25th, 2017}

\begin{abstract}
We demonstrate the non-renormalization of the Chiral Separation Effect
(CSE) in quenched finite-density QCD in both confinement and deconfinement
phases using a recently developed numerical method which allows, for the
first time, to address the transport properties of exactly chiral, dense
lattice fermions. This finding suggests that CSE can be used to fix
renormalization constants for axial current density. Explaining the
suppression of the CSE which we observe for topologically nontrivial gauge
field configurations on small lattices, we also argue that CSE vanishes for
self-dual non-Abelian fields inside instanton cores.
\end{abstract}


\maketitle



Anomalous transport phenomena which involve collective motion of chiral
fermions are important in many disparate sub-fields of physics ranging from
cosmology and
astrophysics~\cite{Vilenkin:80:1,Tashiro:12:1,Metlitski:05:1,Sigl:15:1} over
solid state physics~\cite{Kim:13:1,Kharzeev:14:1,Basar:13:1} to high energy
physics and heavy-ion collision
experiments~\cite{Zhitnitsky:07:1,Kharzeev:08:1,Adamczyk:14:1,Kharzeev:15:2}.
Well-known examples of such phenomena are the induction of vector or axial
currents along the magnetic field in a dense chiral medium, dubbed the Chiral
Magnetic (CME)~\cite{Zhitnitsky:07:1,Kharzeev:08:2} and the Chiral Separation
(CSE)~\cite{Son:04:2,Metlitski:05:1,Son:06:2,Zhitnitsky:07:1} effect,
respectively. In particular, for quark-gluon plasma produced in off-central
heavy-ion collisions, CSE can locally induce large chirality
imbalance~\cite{Kharzeev:11:2}, and, combined with CME, lead to a novel
gapless hydrodynamic excitation - the chiral magnetic wave
(CMW)~\cite{Kharzeev:11:1}.

Within the hydrodynamic approximation, the requirement of positive entropy
production together with the Adler-Bell-Jackiw axial anomaly equation fix
the transport coefficients describing CME and CSE \cite{Son:09:1,Sadofyev:10:1}.
However, the hydrodynamic approximation used in \cite{Son:09:1,Sadofyev:10:1}
might become invalid if the chiral plasma features an infinite correlation length
(e.g. due to spontaneous symmetry breaking~\cite{Buividovich:13:8}), or
interacts with dynamical Yang-Mills fields~\cite{Gursoy:14:1}. This allows
for non-perturbative corrections to CME and CSE. Interactions with dynamic
electromagnetic fields also lead to perturbative corrections
\cite{Jensen:13:1,Miransky:13:1} which we do not consider in this work.

For CSE in QCD matter, which is in the focus of this work, the
non-perturbative correction can be expressed in terms of the in-medium
amplitude $g_{\pi^0 \gamma\gamma}$ of the $\pi^0 \rightarrow \gamma \gamma$
decay~\cite{Son:06:2}:
\begin{equation}
\label{eq:chiralsep} j^5_i = \scsc B_i, \quad \scsc= \scscf \lr{1 - g_{\pi^0 \gamma\gamma}} ,
\end{equation}
where $j^5_i$ is the axial current density and $B_i$ is the external magnetic
field. With $g_{\pi^0 \gamma\gamma} = 0$ we recover the result $\scscf =
\frac{q N_c \mu}{2 \pi^2}$ for $N_c$ species of free chiral fermions (with
$N_c$ being the number of colours), which is also expected to be valid in the
high-temperature phase with restored chiral symmetry~\cite{Metlitski:05:1,Son:06:2,Alekseev:98:01}.

Within the linear sigma model, an estimate of $g_{\pi^0 \gamma\gamma}$ for a
medium with spontaneously broken chiral symmetry and at sufficiently small
quark chemical potential $\mu$ is $g_{\pi^0 \gamma\gamma} = \frac{7
\zeta\lr{3} m^2}{4 \pi^2 T^2}$,
 where $\zeta$ is the Riemann
$\zeta$-function, $m$ is the constituent quark mass and $T$ is the
temperature~\cite{Son:06:2}. With realistic values $m \sim 300 \
\MeV$~\cite{Baboukhadia:97:01} and $T \sim 150 \ \MeV$, which provide a
reasonably good description of the chirally broken phase, we get a correction
of order of $100 \%$ which suppresses the CSE response. Non-perturbative
corrections which suppress CSE were also predicted within the
Nambu-Jona-Lasinio model \cite{Miransky:09:1,Miransky:11:1,Miransky:11:2},
and within the holographic model of a chiral superfluid with broken Abelian
global symmetry \cite{Amado:14:1,Melgar:14:1}.

Since non-perturbative corrections to anomalous transport phenomena might
significantly modify the predictions of anomalous hydrodynamics, it is
important to quantify them in a model-independent way in first-principle
lattice QCD simulations. So far a few lattice studies addressed the infrared
values of anomalous transport coefficients characterizing the CME
\cite{Yamamoto:11:1,Yamamoto:11:2} and the Chiral Vortical Effect (CVE)
\cite{Braguta:13:1,Braguta:14:1}, and found a very significant suppression of
CME and CVE at both low and high temperatures. This is a very puzzling
situation, since at least at high temperatures one can expect that the
thermodynamic consistency arguments \cite{Son:09:1,Sadofyev:10:1} fixing
anomalous transport coefficients in hydrodynamic approximation should be
valid. Possible reasons for this discrepancy might be the use of naively
discretized, non-conserved vector current \cite{Yamamoto:11:1,Yamamoto:11:2}
and energy-momentum tensor \cite{Braguta:13:1,Braguta:14:1}, and the use of
non-chiral Wilson-Dirac fermions in \cite{Yamamoto:11:1,Yamamoto:11:2}. In
summary, this situation clearly calls for more accurate first-principle
studies of anomalous transport coefficients which would be free of systematic
errors.

In this paper we report on a first-principle lattice study of CSE with
finite-density overlap fermions \cite{Bloch:06:01}, which respect the lattice
chiral symmetry at any chemical potential. We use the properly defined
lattice counterpart of the continuum axial current density
$j^5_\mu=\bar{\psi}\gamma_5\gamma_\mu\psi$~\cite{Hasenfratz:02:01,Kikukawa:98:01}:
\begin{equation}
\label{eq:j5}
 \jxa  = \tfrac{1}{2} \bar{\psi} \left( - \gamma_5\K + \K \gamma_5(1-\Dov)\right) \psi ,
\end{equation}
where $\K = \frac{\partial \Dov}{\partial \Theta_{x,\mu}}$ is the derivative
of the overlap operator $\Dov$ over the $U(1)$ lattice gauge field
$\Theta_{x,\mu}$. The lattice axial current \eqref{eq:j5} transforms
covariantly under the lattice chiral symmetry, and is hence protected from
renormalization at zero quark mass and can be directly related to the
continuum axial current in~\eqref{eq:chiralsep}. After some algebra, taking
the expectation value on both sides of equation \eqref{eq:j5} yields
\begin{equation}
  \label{eq:<j5>} \langle \jxa \rangle = \Tr\left(\Dov^{-1}\frac{\partial \Dov}{\partial
\Theta_{x,\mu}} \gamma_5 \right) .
\end{equation}
Technically, the most advanced problem is the calculation of the derivatives
$\frac{\partial \Dov}{\partial \Theta_{x,\mu}}$ which enter the definitions
of conserved vector and axial currents for overlap fermions. To this end we
have developed a special algorithm, described in a separate
paper~\cite{Buividovich:16:2}.

Lattice QCD with dynamical fermions suffers from a fermionic sign problem at
finite quark chemical potential. Moreover, a sign problem seems to be in
general unavoidable for gauge theories with dense fermions in a magnetic field,
since an external magnetic field breaks time-reversal and/or charge-conjugation
symmetries which otherwise ensure the positivity of path integral weight for
gauge theories with iso-spin chemical potential or $SU\lr{2}$ or $G_2$ gauge
groups. To avoid the fermionic sign problem, in the present work we neglect
the effect of sea quarks and work in the quenched approximation, which was
also used for holographic studies of CSE \cite{Amado:14:1,Melgar:14:1}. While
arguments from a QCD random matrix model suggest that in the quenched approximation
any nonzero chemical potential leads to a vanishing chiral condensate and
thus restores chiral symmetry \cite{Stephanov:96:1}, the situation might be
different for a magnetized QCD matter, where random matrix theory becomes
inapplicable, and non-perturbative corrections to CSE appear due to
spontaneous generation of the so-called chiral shift parameter
\cite{Miransky:09:1,Miransky:11:1,Miransky:11:2}, rather than chiral
condensate.

The $SU(3)$ gauge configurations for our calculations were generated using the
tadpole-improved L\"{u}scher--Weisz gauge action~\cite{Luescher:85:1}.  We chose three
different lattice setups: $V=\LT\times \LS^3 = 6\times 18^3$ with $\bzero=8.45$
corresponding to a temperature $T>T_c$ and $V = 14\times 14^3$ and $V = 8 \times 8^3$ with
$\bzero=8.10$ corresponding to $T<T_c$, where $\LT$ and $\LS$ are the temporal and spatial  
extent of the lattice and $T_c \approx 300 \ \MeV$ is the deconfinement transition
temperature of the L\"{u}scher--Weisz action~\cite{Gattringer:02:1}. The physical value of
the lattice spacing $a$ was determined using results from~\cite{Gattringer:01:1}.

For the $14 \times 14^3$ and $6 \times 18^3$ lattices we generated around
$10^3$ gauge configurations, from which we randomly picked $100$ configurations
with topological charge $Q = 0$~\iftoggle{arxiv}{\footnote{One of the configurations for the
parameters $\bzero=8.1$, $\mu = 0.050$ and a magnetic flux of $\MF = 1$
caused a serious breakdown in the Lanczos algorithm when computing the
overlap operator. This could not be fixed and we have only used the remaining
$99$ configurations for this parameter set.}}{\cite{Note1}}. For $6 \times 18^3$ we
also chose $100$ configurations with topological charge $|Q| =  1$, and for $14 \times
14^3$ $111$ with $|Q|= 1$ and $97$ with $|Q|= 2$. For the
$8 \times 8^3$ lattice we generated $5 \cdot 10^3$ configurations, from
which three random sets of $200$ configurations with $Q = 0$, $|Q| =  1$ and
$|Q| =  2$ were selected. We calculated the absolute value of topological
charge $|Q| = |n_R - n_L|$ as the number of zero eigenvalues of the operator
$\Dov \Dov^{\dag}$, relying on the fact that in practice the overlap operator
always has either $n_R = |Q|$ right-handed or $n_L = |Q|$ left-handed zero
modes (see e.g. Sec.~7.3.2 in \cite{GattringerLangLatticeQCD}).

\begin{table}[ht] 
    \begin{tabular}{llrrr}
    \toprule
     \multirow{ 2}{*}{Setup} & $\bzero$ & $8.1$ & $8.1$ & $8.45$ \\
      &  Volume & $14 \times 14^3 $ & $8 \times 8^3 $& $6 \times 18^3 $ \\
     \toprule
    & Lattice & \multicolumn{3}{c}{Phys. Value} \\
    \colrule
    $a~[\fermi]$ & $1$ & $0.125$ & $0.125$ & $0.095$ \\
    $V_S~[\fermi^3]$ & $\LS^3$ & $5.4$ & $1.0$ & $5.0$ \\
    $T~[\MeV]$ & $\LT^{-1}$ & $113$ & $197$ & $346$ \\
    $\mu~[\MeV]$ & $0.050  $ &$79$ & -- & -- \\
    & $0.100  $ & -- & $158$ & -- \\
    & $0.300  $ & $474$ & -- & -- \\
    & $0.040  $ & -- & -- & $83$ \\
    & $0.230  $ & -- & -- & $478$ \\
    $\frac{qB}{\MF}~[\MeV]^2$ & $\frac{2\pi}{a^2\LS^2}$ & $283^2 $ & $495^2$& $289^2 $ \\
\botrule
  \end{tabular}
  \caption{Simulation parameters}
  \label{tab:params}
\end{table}

\begin{figure*}
\subfloat[$T > T_c, \; V=6 \times 18^3$\label{fig:6x18comb}]{%
  \includegraphics[width=0.49\linewidth]{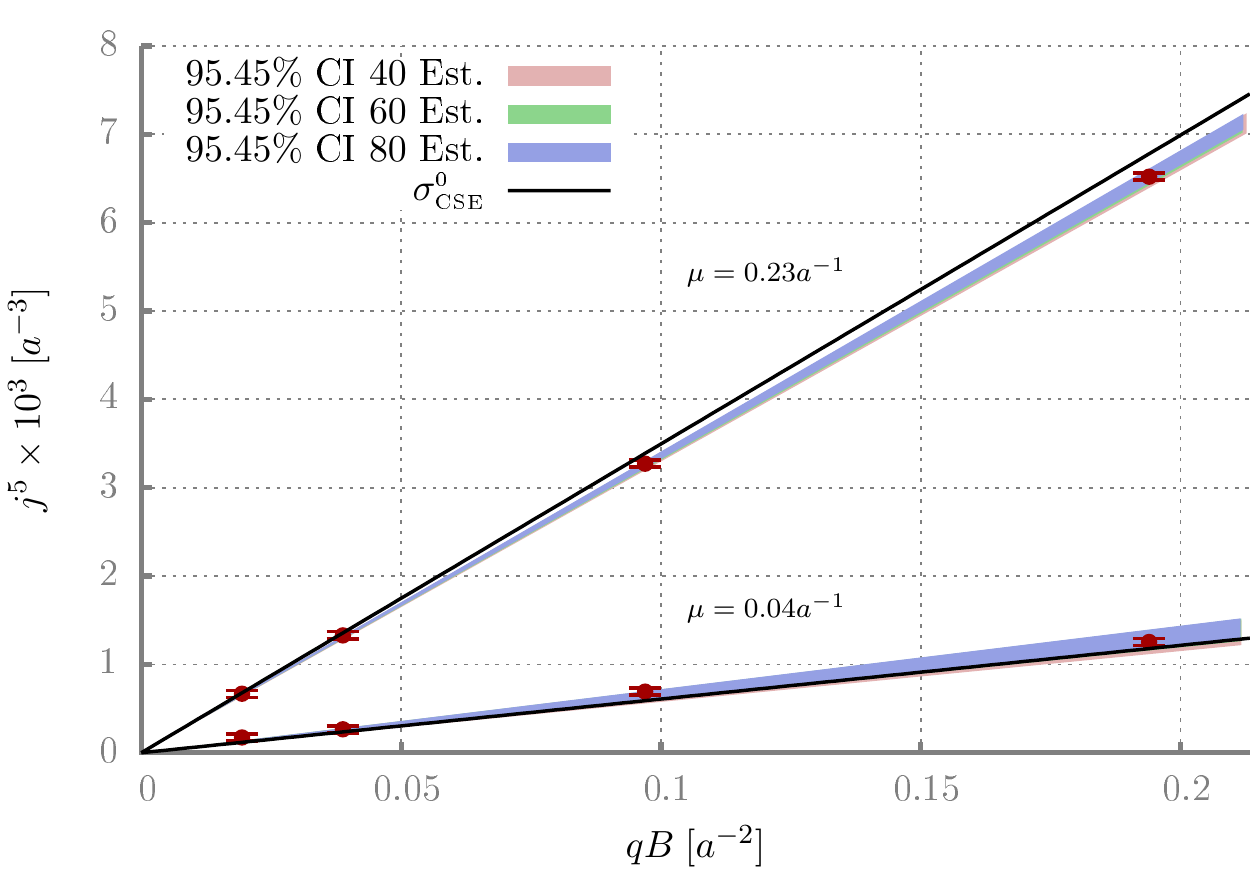}
}\hfill
\subfloat[$T < T_c, \; V=14 \times 14^3$\label{fig:14x14comb}]{%
  \includegraphics[width=0.49\linewidth]{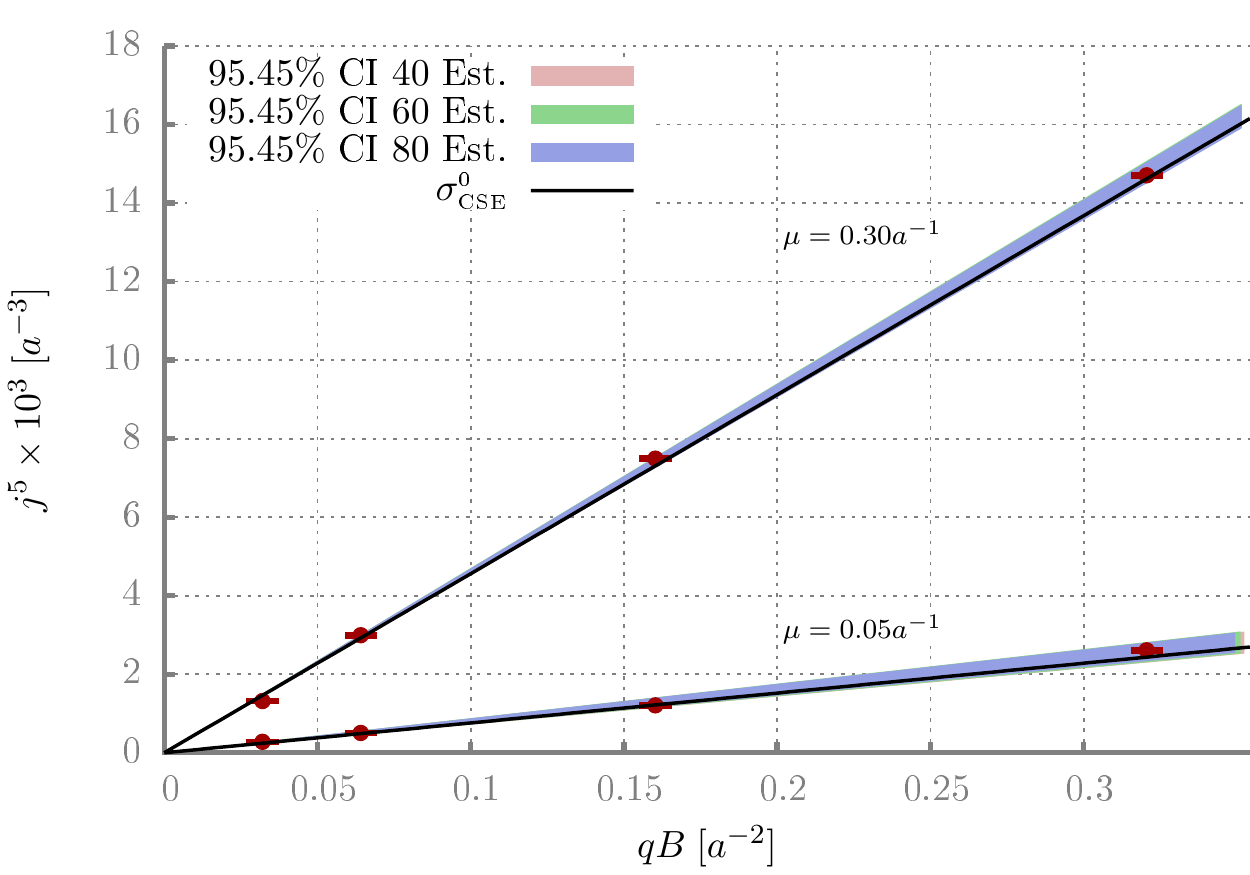}
}
  \caption{Axial current density as a function of the magnetic field strength for
topological charge $Q = 0$ (red circles with error-bars) at $T > T_c$ (on the left) and $T
< T_c$ (on the right). Black lines correspond to the free fermion result $\scscf$ for both
values of the chemical potential. The shaded regions mark the confidence intervals for
$\scsc$ with different numbers of stochastic estimators.}
\end{figure*}

We further introduced a constant, homogeneous external magnetic field following the
prescription of~\cite{Wiese:08:1} with magnetic flux quantum $\MF = 1, 2, 5,10$ for $V = 14 \times 14^3$
and $V = 6 \times 18^3$ at $Q=0$, and $\MF = 0, 1,2,3,4$ for $V = 8 \times 8^3$ at all $Q$.
For $V = 6 \times 18^3$ we chose $\MF = 0, 1,2,3,5$ at $|Q|= 1$ and $\MF = 1,3,5,8,10$ for $V =
14 \times 14^3$ at $|Q|= 1, 2$. For each parameter set, we
evaluated the axial current density averaged over the lattice volume for one or two
different values of the quark chemical potential $\mu$. The trace in equation
\eqref{eq:j5} was calculated using stochastic estimators with $Z_2$-noise. We increased the
number of stochastic estimators until the results were stable (see
Figs.~\ref{fig:6x18comb}~and~\ref{fig:14x14comb} for confidence intervals on $\scsc$ with
different numbers of estimators). For configurations with nonzero topological charge we
introduced a small quark mass $m_q = 0.001 \ a^{-1}$ to make the Dirac operator
invertible. To demonstrate that finite quark mass has practically no effect on $\scsc$,
for the $8 \times 8^3$ lattice we also considered another value $m_q = 0.002 \ a^{-1}$. Our
simulation parameters are summarised in Table~\ref{tab:params}.

The value of $\scsc$ is given by the slope of the axial current density as a function of
the external magnetic field and can be found by performing a one parameter linear fit to
the axial current data (the offset is fixed to zero, since the current has to vanish for
$B=0$). Confidence intervals for $\scsc$ were computed with the statistical bootstrap, by
independently drawing bootstrap samples for every value of $\MF$ and fitting the data
generated in this way.

\begin{figure}[tbh] 
  \includegraphics[width=0.98\linewidth]{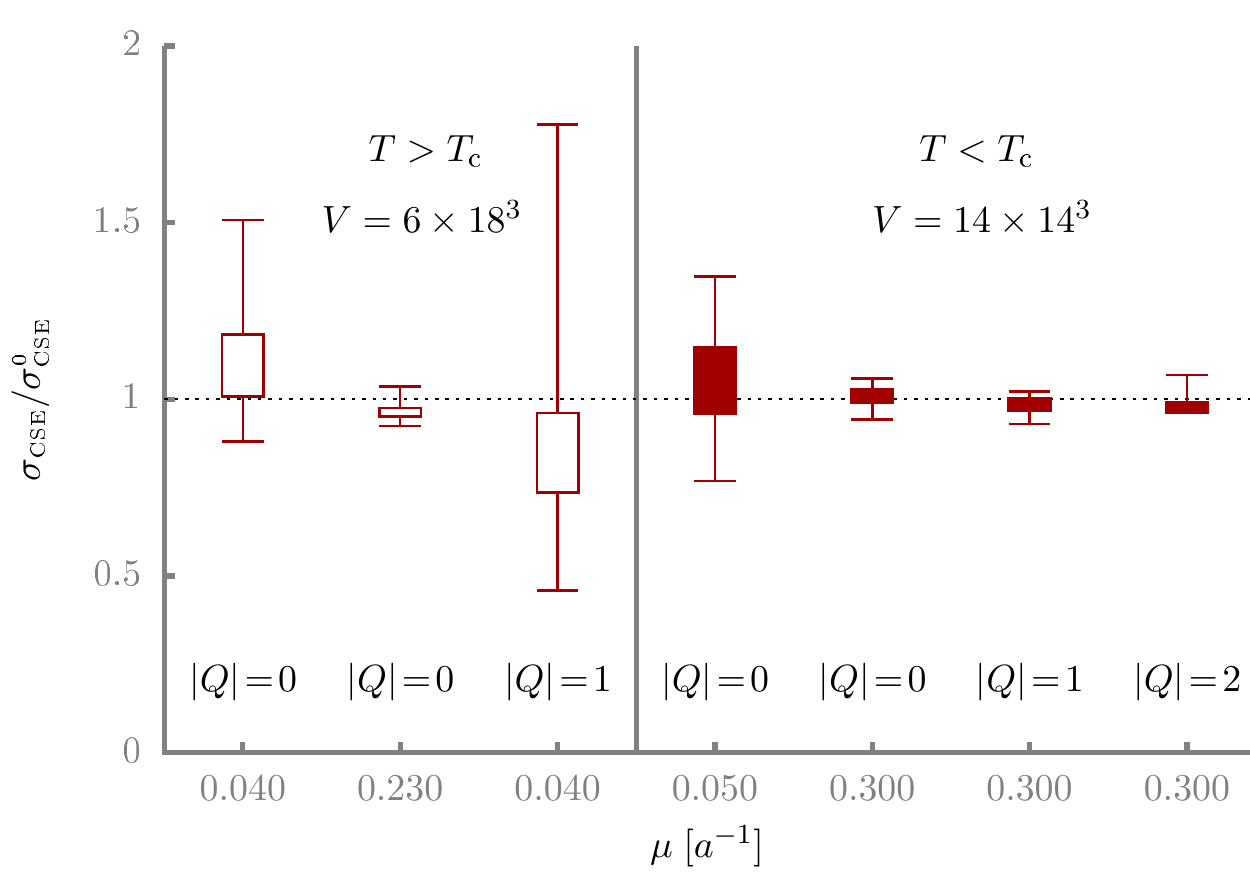}
  \caption{Confidence intervals for the ratio $\scsc/\scscf$ for different values of the
    chemical potential and the topological charge. Boxes and whiskers mark the confidence
    interval for a fit with all data points and with the largest value of $\MF$ excluded,
    respectively. Filled and open boxes are the results for $T < T_c$ and $T > T_c$,
    respectively. }
  \label{fig:CI}
\end{figure}

First we consider the high-temperature deconfinement phase with
$T~=~346~\MeV~>~T_c$ where chiral symmetry should be at least partially
restored as compared to the confinement phase (see e.g.
\cite{Kiskis:01:1,Edwards:00:1} for a discussion of chiral symmetry
restoration in quenched QCD). In this case CSE is expected to have no
corrections to the free fermion
result~\cite{Metlitski:05:1,Alekseev:98:01,Son:06:2}, i.e. $\gpgg\lr{T > T_c}
= 0$. To check this expectation, in Fig.~\ref{fig:6x18comb} we plot our
results for the axial current density for configurations with zero
topological charge as a function of $qB$ (data points with error bars).
Shaded regions show the bootstrap confidence intervals for different numbers
of stochastic estimators which lie on top of each other, hence the error
cannot be improved by using more estimators in the trace calculation. We find
in general a good agreement with the free fermion result $\scscf$, except for
the largest values $\MF=10$ and $\mu = 0.230 \ a^{-1}$, for which we might
see some saturation effect. Therefore we also perform separate fits excluding
the data for $\MF=10$, which show much better agreement with $\scscf$. For
the larger chemical potential value the signal-to-noise ratio is very good
and the relative error of the slope measurement is smaller than $10\%$. For
configurations with $|Q| = 1$ we also find a good agreement with $\scscf$
within statistical errors. The results for the confidence intervals of
$\scsc$ are summarised in Fig.~\ref{fig:CI}. We conclude that within our
statistical errors corrections to CSE are absent in the deconfinement phase
of quenched QCD.

\begin{figure}[t!b!h!]
 \includegraphics[width=0.98\linewidth]{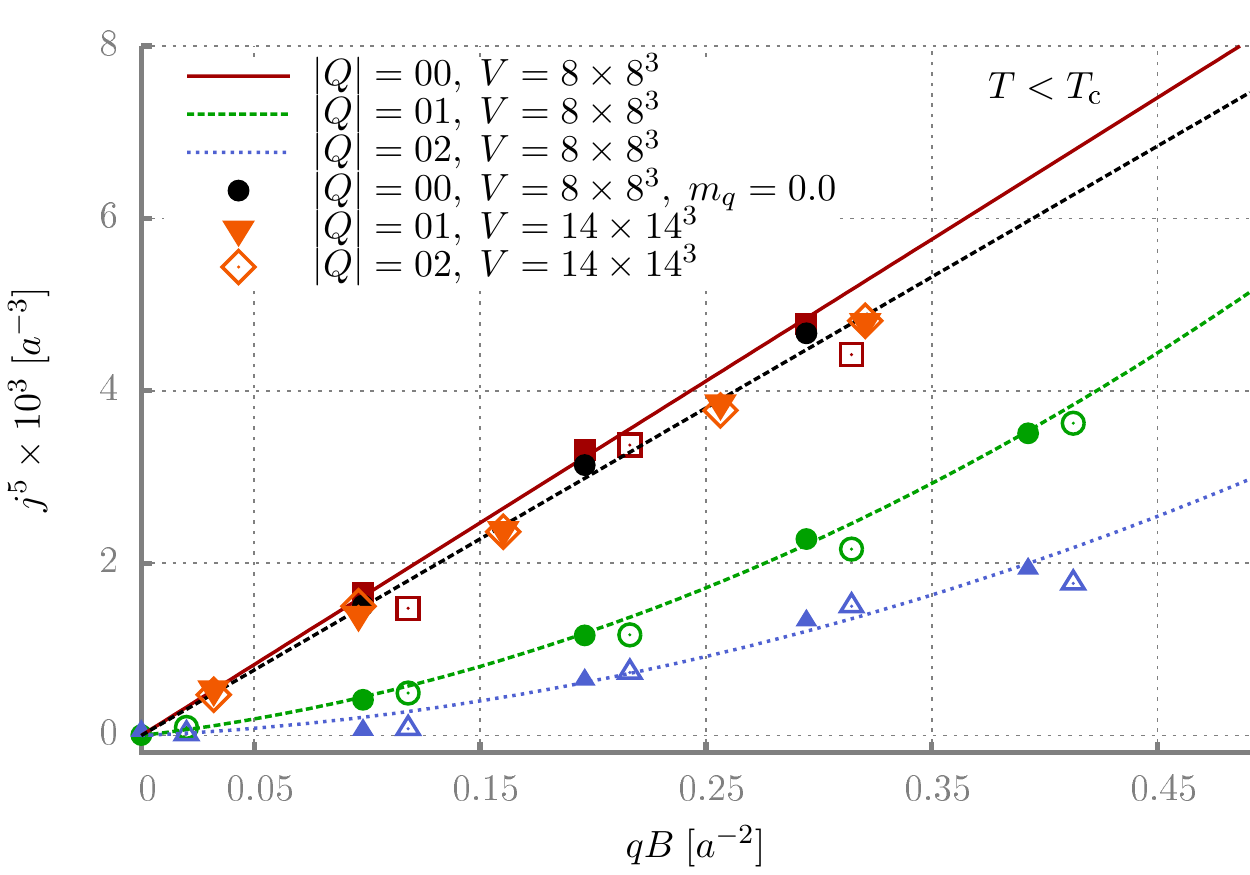}
  \caption{The axial current in different topological sectors. Filled symbols mark the
results for $m_q = 0.001 \ a^{-1}$, the results for $m_q = 0.002 \ a^{-1}$ are shifted by
$0.02 \ a^{-2}$ in the $qB$ axis for better visibility and are marked by open symbols. The
black dots denote the axial current with $Q = 0$ for vanishing quark mass and the black
dashed line corresponds to the free fermion result $\scscf$. To guide the
eye a linear ($Q=0$) or second order polynomial ($|Q|>0$) fit to the data is shown. For
comparison we also plot the results for $|Q|>0$ for the $14 \times 14^3$ lattice.}
  \label{fig:finite_Q}
\end{figure}

We now consider the low-temperature confinement phase at $T < T_c$, where
non-perturbative corrections to CSE can be expected
\cite{Miransky:09:1,Miransky:11:1,Miransky:11:2,Amado:14:1,Melgar:14:1}. In
Fig.~\ref{fig:14x14comb} we plot the axial current density as a function of
the magnetic field strength for gauge field configurations with zero
topological charge on the $14 \times 14^3$ lattice with $T = 113 \ \MeV$ and
$m_q = 0$. The composition of the plot is the same as for
Fig.~\ref{fig:6x18comb}. The confidence intervals for $\scsc$ are very small
and, again, contain the free fermion result within statistical errors. For
the best fits at large chemical potential the relative error of the slope is
smaller than $6\%$. Even for the highest magnetic field strength and the
largest chemical potential we do not see any saturation of the axial current.
For configurations with $|Q| = 1$ we again find that $\scsc$ agrees with the
free fermion result $\scscf$ within confidence intervals (see
Figs.~\ref{fig:CI}~and~\ref{fig:finite_Q}). We thus conclude that even in the
low-temperature phase of quenched QCD the CSE does not receive any
non-perturbative corrections.

At an early stage of this work, we also performed calculations
with small lattice volume $V = 8 \times 8^3$ at $\beta = 8.1$ and $\mu = 0.1
\ a^{-1} = 158 \ \MeV $ (see also Table~\ref{tab:params}). While in the zero
topological sector we found $\scsc$ to agree with the free fermion result
$\scscf$ within statistical errors, which indicates the smallness of
finite-volume effects in the $Q = 0$ sector, for configurations with nonzero
topological charge we found a rather strong suppression of CSE as
well as a non-linear dependence of the axial current on the magnetic field, as
illustrated on Fig.~\ref{fig:finite_Q}. We checked that these findings
are not finite mass effects by doing calculations with two masses $m_q =
0.001 \ a^{-1} = 1.6 \ \MeV$ and $m_q = 0.002 \ a^{-1} = 3.2 \ \MeV$, which yield
almost identical results. Furthermore, we found that the negative
contribution to the axial current which suppresses the CSE comes
exclusively from topological modes of the Dirac operator (the eigenvectors
which would correspond to zero eigenvalues of the Dirac operator with $m_q =
0$.)

In order to understand the possible origin of the suppression of CSE on
topologically nontrivial configurations in small volume, let us consider CSE
in a background of constant Euclidean self-dual non-Abelian gauge field,
which can be interpreted as a limiting case of very large
instanton~\cite{Dunne:11:1}. For gauge configurations with nonzero
topological charge on a small lattice this is a reasonable approximation, as
in most cases only a few instantons would fit in a finite box with physical
size $L = 8 a = 1.0~\fm$ comparable with the characteristic instanton
size~\cite{Shuryak:98:1}. In the gauge where the time-like component of the vector
gauge potential depends only on the longitudinal spatial coordinate, thus
giving rise to a constant chromo-electric field, eigenstates of the Dirac
operator can be labelled by the time-like momentum $k_0$. Introducing a
finite chemical potential leads to a shift $k_0 \rightarrow k_0 - i \mu$.
However, due to relativistic Landau quantisation in an Euclidean electric
field, the dependence of Dirac operator eigenvectors on $k_0$ reduces to
global spatial shifts along the electric field~\cite{Wiese:08:1,Dunne:11:1},
and the corresponding eigenvalues do not depend on $k_0$. Upon analytic
continuation to complex values $k_0 \rightarrow k_0 - i \mu$, this $k_0$
independence translates into the independence of volume-averaged axial and
vector currents on the chemical potential. Since at zero $\mu$ the axial
current vanishes, it also vanishes at finite $\mu$\iftoggle{arxiv}{. We
relegate a more detailed demonstration of this fact to the
\hyperlink{apdx:supplement}{Supplemental Material}.}{~\cite{Note2}.} This argument
suggests a ``porous'' spatial distribution of axial current induced due to CSE, which
should vanish in regions with self-dual gauge fields. Within the instanton liquid model,
these regions can be identified with instanton cores.

To conclude, our numerical study suggests that non-perturbative corrections
to the Chiral Separation Effect are either very small (smaller than our
statistical errors) or vanishing for finite-density quenched QCD in the
thermodynamic limit. By using finite-density overlap fermions
\cite{Bloch:06:1} with covariant axial current we have eliminated systematic
errors due to explicit breaking of chiral symmetry and axial current
renormalization. Finite-volume effects also seem to be rather small at least
in the zero topology sector. Thus the quenched approximation seems to be the only
potentially important source of systematic errors. Indeed, one might argue
that we do not find any non-perturbative corrections predicted in
\cite{Son:06:2}, since quenched QCD at any nonzero chemical potential is in
the chirally symmetric phase with zero chiral condensate
\cite{Stephanov:96:1}. However, the arguments of \cite{Stephanov:96:1} which
are based on random matrix model of QCD might not be directly applicable to
QCD in sufficiently strong external magnetic fields, which should introduce
certain correlations in the otherwise statistically independent entries of
the random matrix which mimics the QCD Dirac operator. Furthermore,
calculations within the Nambu-Jona-Lasinio model
\cite{Miransky:09:1,Miransky:11:1,Miransky:11:2} suggest that
non-perturbative corrections to CSE are related to spontaneous generation of
the so-called chiral shift parameter, which, in contrast to chiral
condensate, cannot be described within the random matrix model framework of
\cite{Stephanov:96:1}. Finally, let us recall that also the holographic
calculations \cite{Amado:14:1,Melgar:14:1} which do predict non-perturbative
corrections to CSE at low temperatures were performed in the quenched
approximation (``probe limit'' in AdS/QCD terminology).

Of course, these arguments simply illustrate that the non-renormalization of
CSE in quenched QCD at both high and low temperatures is a nontrivial result.
They do not prohibit non-perturbative corrections which might originate, for
example, from the complex phase which the fermion determinant acquires at
finite density. Note that external magnetic field renders the fermion
determinant complex-valued even for $SU(2)$ or $G_2$ gauge theories which
are otherwise free of sign problem. Since in massless QCD strong oscillations
of the determinant phase related to the  ``Silver Blaze'' phenomenon can be
expected to set in already at very small density
\cite{Barbour:97:1,Cohen:03:1}, the study of CSE in full QCD with dynamical
fermions might be technically very challenging and is out of the scope of
this work.

The non-renormalization of CSE at least in quenched QCD can also have a
practical application to the calculation of the renormalization constant for
the axial current. Namely, the ratio of the CSE-induced axial current
calculated with non-chiral lattice fermions and/or some non-covariant
discretization of the axial current to the exact result $j^5_i = \frac{\mu}{2
\pi^2} B_i$ yields the multiplicative renormalization constant for the axial
current density in this particular discretization.

Finally we note that the precision with which our results reproduce the
theoretically expected value $\scsc = \scscf$ demonstrates that the approach
to finite-density overlap fermions developed in~\cite{Bloch:06:1} and further
in~\cite{Buividovich:16:2} provides a reliable tool for first-principle
numerical studies of transport properties of dense chiral fermions in lattice
QCD.

\begin{acknowledgments} This work was supported by the S.~Kowalevskaja award from the
Alexander von Humboldt Foundation. The calculations were performed on
``iDataCool'' at Regensburg University, on the ITEP cluster in Moscow and on
the LRZ cluster in Garching. We acknowledge valuable discussions with
G.~Bali, A.~Dromard and A.~Zhitnitsky. MP thanks Rudolf Rödl for helpful
comments on the statistical bootstrap.
\end{acknowledgments}


\iftoggle{arxiv}{
\clearpage
\appendix

\setcounter{equation}{0}
\setcounter{figure}{0}
\setcounter{table}{0}
\setcounter{page}{1}
\makeatletter
\renewcommand{\theequation}{S\arabic{equation}}
\renewcommand{\thefigure}{S\arabic{figure}}

\begin{widetext}

\begin{center}
\textbf{\large Supplemental Material}
\hypertarget{apdx:supplement}{}
\end{center}

\section*{Chiral separation effect in topologically nontrivial background of constant self-dual non-Abelian gauge field}

In order to understand the apparent smallness of the Chiral Separation Effect on topologically nontrivial gauge field configurations, here we consider the simplest topologically nontrivial field configuration with constant and parallel chromo-electric and chromo-magnetic fields of equal strength. If the orientations of both fields in colour space are the same, this configuration is self-dual and has non-zero topological charge proportional to the product of the number of flux quanta of chromo-electric and chromo-magnetic fields. It can be thought of as the instanton of infinitely large size \cite{Dunne:11:1}.

Since the colour orientations of both chromo-electric and chromo-magnetic fields are the same, they can be simultaneously diagonalized in colour space. After that all fermionic observables reduce to sums over $N_c$ Abelian field configurations with parallel electric and magnetic fields, where magnetic fields also include the $U\lr{1}$ Abelian external magnetic field. Thus in order to demonstrate the vanishing of CSE in such a non-Abelian gauge field background, it is enough to show that it vanishes in the background of constant and parallel Abelian electric and magnetic fields $\vec{E} = \vec{e}_3 E$ and $\vec{B} = \vec{e}_3 B$.

Using the gauge where $A_2 = B x_1$ and $A_0 = E x_3$, we can diagonalise the Dirac operator with respect to momenta $k_0$ and $k_2$ conjugate to the time variable $x_0$ and the spatial variable $x_2$. With finite quark mass $m$ and quark chemical potential $\mu$, the Dirac operator in such a background can be represented as 
\begin{eqnarray}
\label{Dirac_operator_EB}
\mathcal{D}= \gamma_{\mu} \nabla_{\mu} + \mu \gamma_0 + m = 
\left(
  \begin{array}{cc}
      m     & -i W^- \\
    - i W^+ &     m \\
  \end{array}
\right),
 \quad
W^{\pm} = \left(
  \begin{array}{cc}
    w_E a_E^{\pm} & \pm w_B a_B^{\dag} \\
    \mp w_B a_B & w_E a_E^{\mp}  \\
  \end{array}
\right),
\nonumber \\
 a_{E}^{\pm} = \frac{\partial_3}{w_E} \mp \frac{w_E}{2} \, \lr{x_3 - \frac{k_0 - i \mu}{E}},
 \quad 
 a_{B} = \frac{\partial_1}{w_B} + \frac{w_B}{2} \, \lr{x_1 - \frac{k_2}{B}},
 \quad
 w_E = \sqrt{2 E}, \quad w_B = \sqrt{2 B},
\end{eqnarray}
where we have introduced the creation/annihilation operators $a_E^{\pm}$ and $a_B^{\dag}$, $a_B$ which describe the relativistic Landau quantisation of fermion motion in constant external fields in $\lr{x_0, x_3}$ and $\lr{x_1, x_2}$ planes. These operators satisfy the usual bosonic commutation relations $\lrs{a_B, a_B^{\dag}} = 1$ and $\lrs{a_E^-, a_E^+} = 1$. An interesting new feature is that at finite chemical potential $\mu$ the creation and annihilation operators $a_E^+$ and $a_E^-$ are no longer Hermite conjugate of each other. The reason is that, as one can see from (\ref{Dirac_operator_EB}), finite chemical potential results in a complex-valued shift of the potential minimum of a ``harmonic oscillator'' described by the non-Hermitian ``Hamiltonian'' $a_E^+ a_E^-$. As a consequence, also the $\gamma_5$-hermiticity of the Dirac operator is lost at finite chemical potential.

The propagator corresponding to the Dirac operator (\ref{Dirac_operator_EB}) can be expressed as
\begin{eqnarray}
\label{Dirac_propagator_EB}
 \mathcal{D}^{-1} = \left(
  \begin{array}{cc}
      m \lr{m^2 + W^+ W^-}^{-1} & i \lr{m^2 + W^+ W^-}^{-1} W^+ \\
     i W^- \lr{m^2 + W^+ W^-}^{-1} &     \lr{m^2 + W^- W^+}^{-1} \\
  \end{array}
\right) ,
 \nonumber \\
 W^+ W^- = \left(
  \begin{array}{cc}
      w_E^2 n_E + w_B^2 n_B & 0 \\
      0 &     w_E^2 n_E + w_B^2 n_B + w_E^2 + w_B^2 \\
  \end{array}
\right) ,
\end{eqnarray}
where we have introduced the ``electric'' and ''magnetic'' number operators $n_E = a_E^+ a_E^-$ and $n_B = a_B^{\dag} a_B$, whose eigenvalues label Landau levels in the $\lr{x_0, x_3}$ and $\lr{x_1, x_2}$ planes. This construction is very similar to the one used in~\cite{Dunne:11:1}, with the only difference that $a_E^+$ and $a_E^-$ are no longer conjugate. 

The expectation value of axial current can be now expressed as 
\begin{eqnarray}
\label{JA_VEV_EB}
 J_{A 3}
 = \tr\lr{\mathcal{D}^{-1} \gamma_5 \gamma_3} 
 = i \tr\lr{\lr{W^+ \sigma_3 + \sigma_3 W^-} \lr{m^2 + W^+ W^-}^{-1}} 
 = \nonumber \\ =
 i w_E \tr\lr{a_E^+ + a_E^-} \lr{G\lr{n_E, n_B} - G\lr{n_E+1,n_B+1}} ,
\end{eqnarray}
where we have introduced the notation $G\lr{n_E, n_B} = \frac{1}{m^2 + w_E^2 n_E + w_B^2 n_B}$.

The simplest way to proceed now is to express the trace in the above equation in the basis of direct product of eigenstates of bosonic number operators $n_E$ and $n_B$. While the eigenstates of $n_B$ are the usual harmonic oscillator wave functions corresponding to Landau levels in constant magnetic field, eigenstates of the operator $n_E = a_E^+ a_E^-$ which is non-Hermitian at finite $\mu$ deserve a more detailed description. Namely, we first define the left and right ``ground state'' eigenvectors $\bra{L_0}$ and $\ket{R_0}$ of $a_E^+ a_E^-$ which satisfy the equations
\begin{eqnarray}
\label{LR_lowest_eigenvectors}
 a_E^- \ket{R_0} = 0, \quad \bra{L_0} a_E^+ = 0, \quad \braket{L_0}{R_0} = \int\limits_{-\infty}^{+\infty} dx_3 \braket{L_0}{x_3} \braket{x_3}{R_0} = 1 ,
 \nonumber \\
 \braket{x_3}{R_0} = \lr{E/\pi}^{1/4} \expa{-\frac{E}{2} \lr{x_3 - \frac{k_0 - i \mu}{E}}^2 },
 \quad
 \braket{L_0}{x_3} = \lr{E/\pi}^{1/4} \expa{-\frac{E}{2} \lr{x_3 - \frac{k_0 - i \mu}{E}}^2 } .
\end{eqnarray}
We can now define the left and right ``excited state'' eigenvectors by successively multiplying $\ket{R_0}$ by $a_E^+$ and $\bra{L_0}$ by $a_E^-$:
\begin{eqnarray}
\label{LR_highest_eigenvectors}
 \ket{R_n} = \frac{\lr{a_E^+}^n}{\sqrt{n!}} \ket{R_0}, 
 \quad
 \bra{L_n} = \bra{L_0} \frac{\lr{a_E^-}^n}{\sqrt{n!}}, 
\end{eqnarray}
Using the identity $a_E^- \lr{a_E^+}^n = \lr{a_E^+}^n a_E^- + n \lr{a_E^+}^{n-1}$ which follows from the commutation relations $\lrs{a_E^-, a_E^+} = 1$ and the normalization condition $\braket{L_0}{R_0} = 1$, it is easy to check that $\ket{R_n}$ and $\bra{L_n}$ form the full bi-orthogonal eigenbasis of the operator $a_E^+ a_E^-$, with the eigenvalues $n_E$ corresponding to $\ket{R_n}$ and $\bra{L_n}$. It is easy to check that the ``wave functions'' $\braket{x_3}{R_n}$ and $\braket{L_n}{x_3}$ are equal to each other and are simply analytic continuation of the conventional real-valued harmonic oscillator wave functions to the complex-valued position $\lr{k_0 - i \mu}/E$ of the potential minimum. On the other hand, exactly because $\braket{x_3}{R_0}$ and $\braket{L_0}{x_3}$ are now complex-valued, $\ket{R_n}$ and $\bra{L_n}$ are not complex conjugates of each other.

Using the eigenbasis decomposition of $n_B$ and $n_E$, we can now rewrite the volume-averaged axial current (\ref{JA_VEV_EB}) as
\begin{eqnarray}
\label{axial_current_EB_s3}
J_{A \, 3} = i w_E \sum\limits_{n_E, n_B} \bra{L_n} \lr{a_E^+ + a_E^-} \ket{R_n} \lr{G\lr{n_E, n_B} - G\lr{n_E+1, n_B+1}}
 = 0 .
\end{eqnarray}
We conclude that the axial current vanishes due to the fact that the matrix element $\bra{L_n} \lr{a_E^+ + a_E^-} \ket{R_n} = \sqrt{n_E+1}\braket{L_n}{R_{n+1}} + \sqrt{n_E}\braket{L_n}{R_{n-1}}$ is zero for any $n$ by virtue of the bi-orthogonality relation $\braket{L_n}{R_m} = \delta_{nm}$. Hence CSE vanishes for such self-dual topologically nontrivial backgrounds of constant non-Abelian gauge fields.

\end{widetext}
}{}


\end{document}